\documentclass[11pt]{article}

\usepackage{bbm}
\usepackage{amsfonts}
\usepackage{amsmath}
\usepackage{amssymb}
\usepackage{amsthm}
\usepackage[mathscr]{eucal}
\usepackage{bbold}
\usepackage{braket}
\usepackage{color}
\usepackage{dsfont}
\usepackage{framed}
\usepackage{jheppub}
\usepackage{mathtools}
\usepackage{physics}
\usepackage[normalem]{ulem}
\usepackage{tensor}
\usepackage{thmtools}
\usepackage{thm-restate}
\usepackage{ulem}
\usepackage{tikz}
\usepackage{soul}

\usetikzlibrary{math} 

\newcommand{\ignore}[1]{}

\renewcommand{\(}{\left(}
\renewcommand{\)}{\right)}
\renewcommand{\[}{\left[}
\renewcommand{\]}{\right]}

\makeatletter\def\@fpheader{~}\makeatother
\newcommand{\Eqref}[1]{Eq.~\eqref{#1}}
\newcommand{\secref}[1]{Sec.~(\ref{#1})}
\newcommand{\figref}[1]{Fig.~(\ref{#1})}

\definecolor{bcolor}{RGB}{10,10,255}

\usepackage[export]{adjustbox}
 \usepackage{simpler-wick}

\title{Computable Cross Norm in Tensor Networks and Holography}

\author{Alexey Milekhin,}
\emailAdd{milekhin@ucsb.edu}

\author{Pratik Rath}
\emailAdd{rath@ucsb.edu}

\author{and Wayne Weng}
\emailAdd{wweng@physics.ucsb.edu}

\affiliation{Department of Physics, University of California, Santa Barbara, CA 93106, USA}

\abstract{The Computable Cross Norm (CCNR) was recently discussed in Ref.~\cite{Yin:2022toc} as a measure of multipartite entanglement in a condensed matter context. In this short note, we point out that it is closely related to the $(2,n)$-R\'enyi reflected entropy, which has been studied in the context of AdS/CFT. We discuss the calculation of the CCNR in random tensor networks as well as holographic CFTs. The holographic dual involves a backreacted entanglement wedge cross section in a geometry sourced by R\'enyi-2 cosmic branes. We perform explicit calculations for two intervals in a hyperbolic random tensor network as well the vacuum state of a 2D holographic CFT, and analyze the occurence of a connected-to-disconnected phase transition. The example illustrates the validity of the proposal for analytic continuation in holography for arbitrary values of R\'enyi parameter $n$. We comment on a symmetry-resolved generalization of this quantity. 
}

\begin{document}
 
 
\maketitle

\section{Introduction}\label{sec:intro}
Recently, Ref.~\cite{Yin:2022toc} discussed the so-called \textit{computable cross-norm} (CCNR) as a measure of entanglement for bipartite mixed states. Given a density matrix $\rho_{AB}$ on subsystems $A$ and $B$, the realignment matrix $R$ is defined as \cite{2002quant.ph..5017C,2003JPhA...36.5825R}
\begin{equation}
    \bra{a}  \bra{a'}  R  \ket{b} \ket {b'} := \bra{a} \bra{b} \rho \ket{a'} \ket{b'}. 
\end{equation}
In general, this is a rectangular matrix of dimensions $d_A^2 \times d_B^2$, $d_{A,B}$ being the corresponding Hilbert space dimensions. Then, the CCNR is defined as
\begin{equation}\label{eq:ccnr}
    || R ||_1 = \Tr \( \sqrt{R R^\dagger} \).
\end{equation}
The usefulness of the CCNR as an entanglement measure arises from the fact that if $\rho$ is separable, then $||R||_1 \le 1$.\footnote{The converse is not true.} More generally, one can define a R\'enyi version of the CCNR as
\begin{equation}\label{eq:renyi}
    Z_n = \Tr \( R R^\dagger \)^n.
\end{equation}

Now, a similar construction appeared in the context of AdS/CFT under the guise of reflected entropy \cite{Dutta:2019gen}. The reflected entropy is a von Neumann entropy computed in a specific purification of $\rho_{AB}$ called the canonical purification. The canonical purification of $\rho_{AB}$ is defined as the state $\ket{\sqrt{\rho_{AB}}}$, where we compute the matrix square-root and then interpret it as a state in the vector space of operators acting on $A\cup B$. This state is a purification of the original density matrix and lives on a doubled Hilbert space consisting of $A\cup B$ and their mirror images $A^*\cup B^*$. A R\'enyi generalization of the canonical purification is $\ket{\rho_{AB}^{m/2}}$, which is useful since it can be constructed using a replica trick for even integer $m$. Given the state $\ket{\rho_{AB}^{m/2}}$, we can trace out $BB^*$ to obtain the density matrix $\rho_{AA^*}^{(m)}$, whose R\'enyi entropy can then be computed:
\begin{equation}\label{eq:SRmn}
    S_R^{(m,n)}=\frac{1}{1-n}\ln  \[\Tr\(\rho_{AA^*}^{(m)}\)^n\].
\end{equation}
The above quantity is referred to as the $(m,n)$-R\'enyi reflected entropy.

Now, the key observation is that $\rho_{AA^*}^{(m)}$ for $m=2$ is the same as $ R R^\dagger$ in \Eqref{eq:ccnr}. Thus, the quantity defined in \Eqref{eq:renyi} is identical to that inside the square brackets in \Eqref{eq:SRmn}, after we set $m=2$. Based on this, we will import some known results from the AdS/CFT literature and apply it to the CCNR. Moreover, we also discuss some novel results for the $(2,n)$-R\'enyi reflected entropy in random tensor networks (RTNs) and AdS/CFT.

We first analyze this problem in RTNs in \secref{sec:RTN} and then for AdS/CFT in \secref{sec:hol}. As noted previously in Ref.~\cite{Akers:2021pvd}, the saddles includes cosmic branes located at the entanglement wedge cross section. Moreover, for general $n>1$, they have been noted to backreact and squeeze the cross section to a smaller size. Here, we will additionally discuss the case of $n<1$, where they instead expand to a larger size due to the negative tension. For holographic systems, we propose an analytic continuation in the parameter $n$, a la Lewkowycz-Maldacena \cite{Lewkowycz:2013nqa}. Moreover, Ref.~\cite{Yin:2022toc} computed $Z_n$ for two intervals in the vacuum state of a 2D CFT and showed that it is proportional to the torus partition function.\footnote{For non-holographic CFTs, calculations of the reflected entropy (at $m=1$) were done in Refs.~\cite{Bueno:2020fle,Dutta:2022kge}.} Using this, we analytically compute $Z_n$ and demonstrate the validity of our proposed analytic continuation for arbitrary values of $n$. The connected-to-disconnected entanglement wedge transition in this quantity is directly related to the Hawking-Page transition.

\section{Random Tensor Networks}
\label{sec:RTN}

RTNs are toy models of holography, defined in Ref.~\cite{Hayden:2016cfa}, which have served useful for computations of various multipartite entanglement measures like the reflected entropy \cite{Akers:2021pvd,Akers:2022zxr,new} and entanglement negativity \cite{2021JHEP...06..024D,2022JHEP...02..076K,negativity}.\footnote{Another quantity that measures multipartite entanglement was recently studied in Refs.~\cite{Gadde:2022cqi,Penington:2022dhr}.} The replica trick to compute $Z_n$ involves taking $2n$ copies of the original state and applying the following permutations on the boundary subregions: $g_A=(1\, 2)(3 \,4)\dots (2n-1\, 2n)$ on $A$, $g_B=(2\, 3)(4\, 5)\dots (2n\, 1)$ on $B$ and $e$ (identity) on $C$. The computation of $Z_n$ boils down to the calculation of the partition function of a spin model with degrees of freedom in the permutation group $\text{Sym}_{2n}$. Energy costs for the model are defined by the Cayley distance function $d(g_1,g_2)$ which measures the number of transpositions required to go from $g_1$ to $g_2$. One can check that $d(g_A,e)=d(g_B,e)=n$ and $d(g_A,g_B)=2(n-1)$.

\begin{figure}
    \centering
   \minipage{0.3\textwidth}
   \includegraphics[scale=0.6]{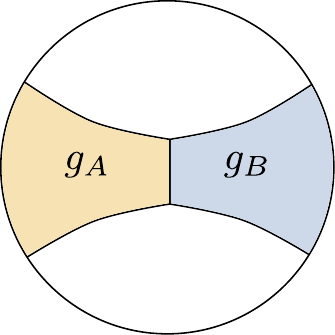}
   \endminipage
      \minipage{0.3\textwidth}
   \includegraphics[scale=0.6]{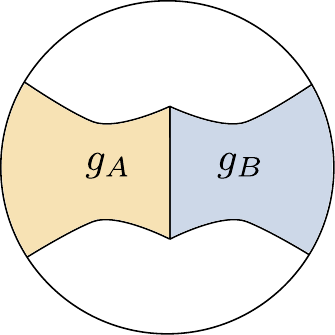}
   \endminipage
      \minipage{0.3\textwidth}
   \includegraphics[scale=0.9]{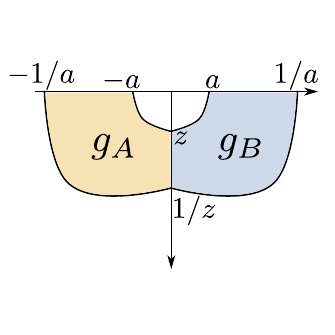}
   \endminipage
   
    \caption{The ground states of the spin-model partition function that computes $Z_n$ in the RTN in the connected phase. In a hyperbolic RTN, the domain wall separating the $g_A$ and $g_B$ domains is a squeezed (expanded) EWCS for $n>1$ ($n<1$). The last plot shows the geometry in Poincar\'e coordinates.}
    \label{fig:EWCS}
\end{figure}

In the large bond dimension limit (or low temperature limit), the partition function is dominated by its ground state. It can be proved that the minimal energy configuration only consists of domains of the elements $e$, $g_A$ and $g_B$ \cite{new}.\footnote{For $m\neq2$, the element $X$ discussed in Refs.~\cite{Akers:2021pvd} is also important. But at $m=2$, $X=e$ and thus, we do not need to consider it separately.} The solutions are completely determined by the tensions of the domain walls, which can be analytically continued to non-integer $n$ in an obvious manner. The solutions thus take the form shown in \figref{fig:EWCS}. The ``entanglement wedge cross section" (EWCS) domain wall separating $g_A$ and $g_B$ squeezes in the entanglement wedge determined by the domain wall separating $g_A,g_B$ and $e$ when $n>1$. As is familiar for the usual reflected entropy, the tension vanishes in the limit $n\to 1$ resulting in the usual EWCS. Finally, we will also be interested in $n<1$, where the cross section expands out the entanglement wedge as shown in \figref{fig:EWCS}.

\begin{figure}
    \centering
    \includegraphics{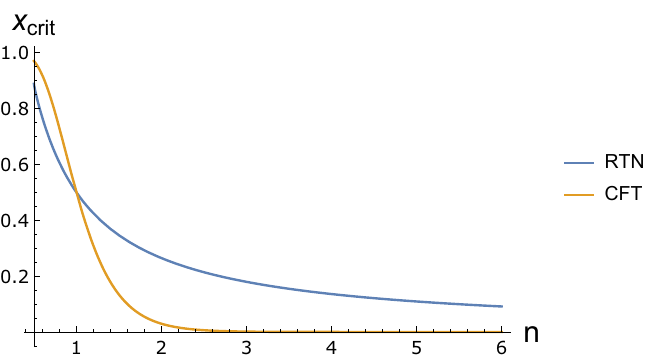}
    \caption{Critical cross-ratio $x$ as a function of R\'enyi index $n$ for hyperbolic RTN (blue) and holographic CFT (orange). For RTN, the answer for $n<1/2$ formally becomes complex, so we only plot $n\geq 1/2$.} 
    \label{fig:phase}
\end{figure}

As an illustrative example, we compute $Z_n$ in hyperbolic networks. This is identical to the calculation done in Ref.~\cite{Akers:2021pvd}, except that we also analyze $n<1$. Working in Poincar\'e hyperbolic coordinates, we label the points as shown in \figref{fig:EWCS} (borrowed from Ref.~\cite{Akers:2021pvd}). Following the notation of Ref.~\cite{Akers:2021pvd}, $Z_n$ is given by $\min\(\exp\[f_\text{II}\],\exp\[f_\text{IV}\]\)$, where
\begin{align}
    f_\text{II} &= \max_{z\geq 0}\, \exp\[\ln \chi \(-4(n-1)\ln z + 4n\ln \frac{1}{2}\(\frac{a}{z}+\frac{z}{a}\)\)\]\\
    f_\text{IV} &= 2n \ln \(\frac{1-x}{x}\)
\end{align}
where $\chi$ is the bond dimension in the network, $a$ is related to the size of the interval (see \figref{fig:EWCS}),  and $x$ is the two-interval cross-ratio. $f_\text{II}$ corresponds to the connected phase, whereas $f_\text{IV}$ corresponds to the disconnected phase. After performing the maximum for $f_\text{II}$, we obtain
\begin{equation}
    f_\text{II} = 4(n-1) \log \frac{1+\sqrt{1-x}}{\sqrt{x}} + 4n \( H \( \frac{1}{2n},1-\frac{1}{2n} \)-\log 2 \),
\end{equation}
where
\begin{equation}
    H(p_+,p_-) = - p_+ \log p_+ - p_- \log p_-.
\end{equation}
The location of the phase transition is determined by the condition $f_\text{II}=f_\text{IV}$. The location is $n$-dependent, unlike the case of R\'enyi entropy for two intervals \cite{2010PhRvD..82l6010H}, and we plot this critical cross ratio as a function of $n$ in \figref{fig:phase} (blue). For all $x$, $Z_{1/2}$ (which is the CCNR) stays below $1$, so formally we can not conclude anything about separability. We note however that this is consistent with the fact that in the disconnected phase we expect the state to be separable.

The maximum for $f_\text{II}$ is achieved at $z=\sqrt{2n-1}\,a$. This demonstrates clearly that the EWCS is squeezed for $n>1$ and expanded for $n<1$. Interestingly, the EWCS goes all the way to the boundary at $n=\frac{1}{2}$, which is relevant to computing the CCNR. It is unclear if this analytic continuation fails beyond that point. It would be interesting to provide a more rigorous derivation of this analytic continuation beyond $n=1$, since there are other known situations where negative tension branes lead to large corrections to the naive saddle \cite{Akers:2022max}. Thus, the analytic continuation may be subtle, although we see no obvious reason for it to fail for $n>\frac{1}{2}$.

\section{AdS/CFT}
\label{sec:hol}

\begin{figure}
    \centering
    \includegraphics[scale=0.1]{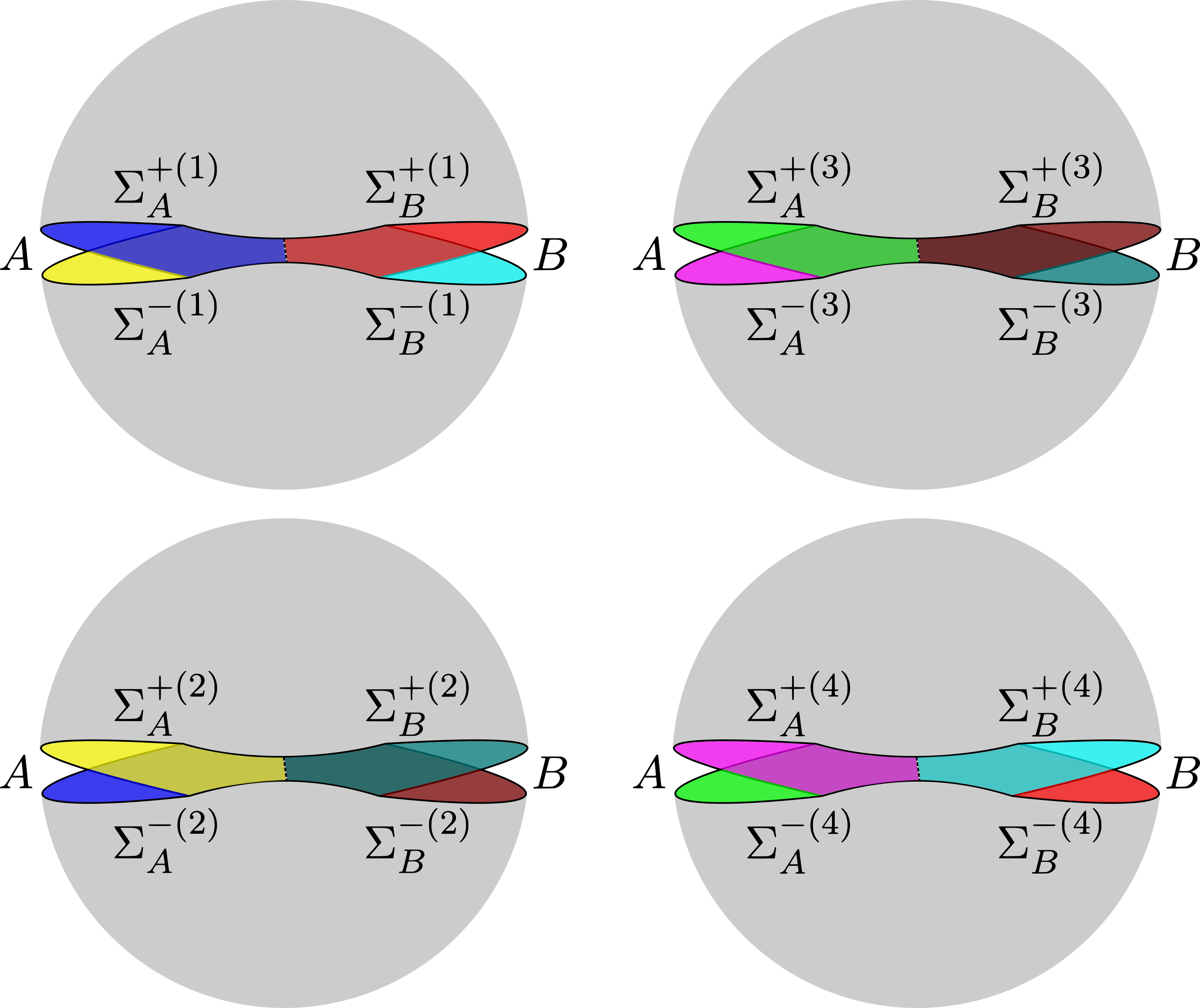}
    \caption{The replica partition function $Z_n$ is computed by the above gluing procedure on the boundary. Each boundary subregion has a bulk domain homologous to it which is replicated in the same topological fashion (represented by the different colours). Here we depict the case $n = 2$ as an example.}
    \label{fig:replica}
\end{figure}

Now, we consider the same problem in holography. The replica partition function $Z_n$ sets boundary conditions for the gravitational path integral (see \figref{fig:replica}). The boundary conditions have a $\mathbb{Z}_n$ symmetry under the permutation $(1 \,3\, 5 \dots 2n-1)(2 \,4\, 6\dots 2n)$. Assuming that the bulk saddle $B_{2n}$ also inherits this $\mathbb{Z}_n$ symmetry, we can quotient the bulk solution to obtain an orbifolded spacetime $\hat{B}_{2n}=B_{2n}/\mathbb{Z}_n$ following Ref.~\cite{Lewkowycz:2013nqa}. The fixed points of this $\mathbb{Z}_n$ symmetry then become conical defects with opening angle $\frac{2\pi}{n}$. 

\begin{figure}
    \centering
    \includegraphics[scale=0.1]{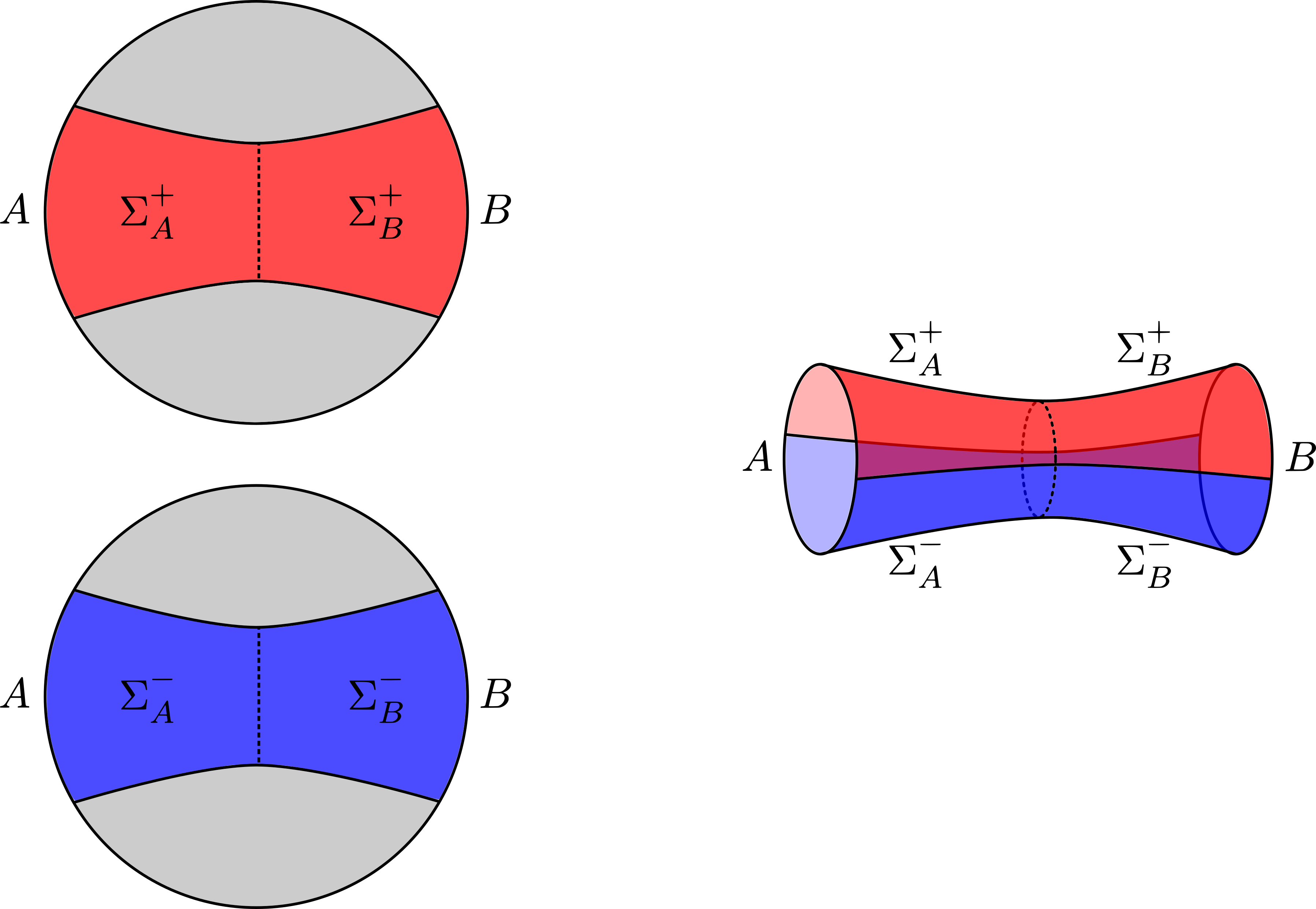}
    \caption{The Cauchy slice of $\hat{B}_{2n}$ consists of two copies of the backreacted entanglement wedge (left) glued together to form a cylinder (right). The fixed points of the $\mathbb{Z}_n$ symmetry lie on the throat of the cylinder and correspond to two copies of the squeezed/expanded EWCS.}
    \label{fig:cyl}
\end{figure}

For simplicity, let us consider a geometry with a time symmetric Cauchy slice. Then the structure of the conical defects takes the form shown in \figref{fig:cyl}. This is essentially the spatial geometry of the R\'enyi generalization of the canonical purification. Note that $\hat{B}_{2n}$ contains two copies of subregions $A$ and $B$, the additional copies being $A^*$ and $B^*$ as discussed before. Moreover, the geometry is not the same as the original spacetime, since it is a double cover that includes backreaction. Thus, when $n\to1$, the geometry is related to the spacetime used in computing the 2nd R\'enyi entropy (this is essentially $m=2$ in terms of the $(m,n)$-R\'enyi reflected entropy).

Assuming that this replica symmetric solution dominates, we obtain a bulk picture that is now analytic in $n$, and can be continued away from integer $n$. We again obtain a squeezed EWCS for $n>1$ and an expanded EWCS for $n<1$. However, the calculations here are different from the RTN due to the fact that the cosmic branes backreact on the geometry.

\begin{figure}
    \centering
    \includegraphics[scale=0.1]{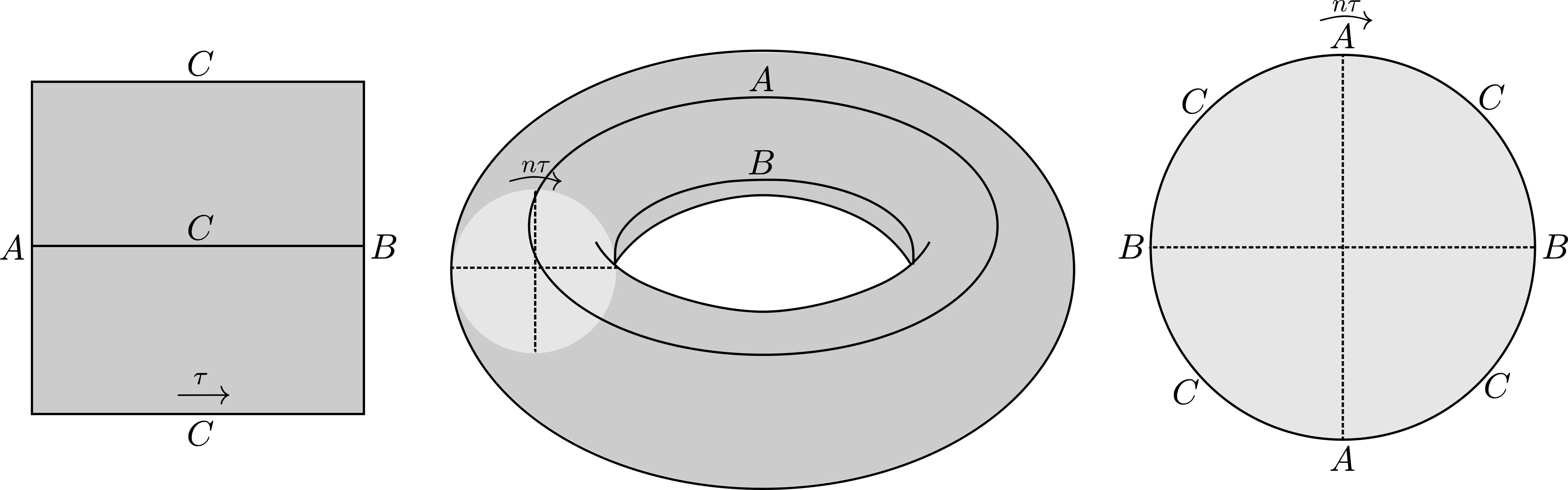}
    \caption{(Left) A single copy of the asymptotic boundary, mapped to a flat cylinder. The top and bottom are periodically identified. (Middle) Gluing $n$ copies of the flat cylinder along the $\tau$ direction forms a torus with modular parameter $n \tau$. In the connected phase, the $n\tau$ direction is chosen to be the contractible cycle in the bulk. This corresponds to a BTZ black hole with $\beta = 2\pi n$. (Right) A cross-section of the torus.}
    \label{fig:torus}
\end{figure}

\begin{figure}
    \centering
    \includegraphics[scale=0.1]{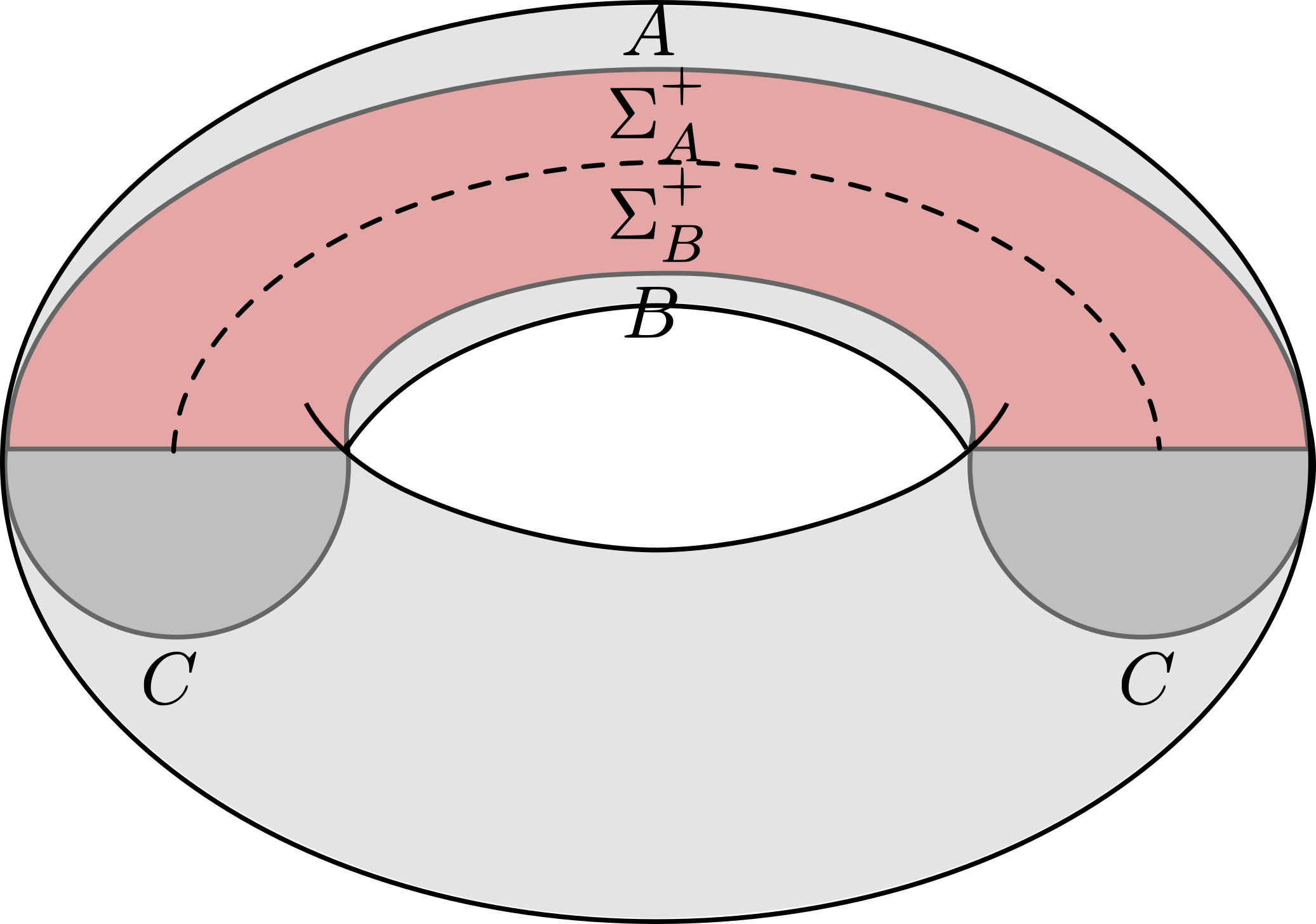}
    \caption{A slice of the $\mathbb{Z}_n$-quotiented BTZ corresponding to the Cauchy slice of $\hat{B}_{2n}$. The dotted line is a conical defect at the would-be horizon, with opening angle $2\pi/n$. When $n > 1$, the defect is a deficit and the EWCS is squeezed. When $n < 1$, the defect is an excess and the EWCS is expanded.}
    \label{fig:torus_slice}
\end{figure}

Now for the case of two intervals in 2D CFT, Ref.~\cite{Yin:2022toc} showed that the $n$-R\'enyi CCNR is proportional to the torus partition function with modulus $n \tau$\footnote{For details on the conformal mapping, see Refs.~\cite{Lunin:2002fch,Headrick:2010zt,Antonini:2022sfm}.}, i.e.,
\begin{equation}
   Z_n = \frac{1}{k^{n c/12}} Z_\text{CFT}(n \tau), 
\end{equation}
\begin{equation}
    k = l_a l_b |u_a - u_b| |v_a-v_b| |u_a-v_b||u_b-v_a|,
\end{equation}
where intervals $A=[u_a,v_a]$, $B=[u_b,v_b]$ have lengths $l_a=|u_a-v_a|$, $l_b=|u_b-v_b|$. The corresponding cross-ratio $x$ is\footnote{Note that this definition of the cross-ratio is related to the one in Ref.~\cite{Yin:2022toc} by $x_{\text{here}} =  1-x_{\text{there}}$.}
\begin{equation}
    1-x = \frac{(u_a-v_a)(u_b-v_b)}{(u_a-u_b)(v_a-v_b)}.
\end{equation}
Purely imaginary $\tau$ is determined in terms of $x$ as
\begin{equation}
   1 -x = \left( \frac{\theta_2(\tau)}{\theta_3(\tau)} \right)^4,
\end{equation}
where $\theta_y(\tau)$ are Jacobi theta-functions (\verb|EllipticTheta[y,e^{i \pi \tau}]| in \verb|Mathematica|). 
In general, the torus partition function for CFTs is not a universal quantity. However, for holographic CFTs the leading large-$c$ answer is known. When $|\Re \tau| \leq \Im \tau$, the corresponding three-dimensional geometry is either thermal AdS$_3$ or a Ba\~nados--Teitelboim--Zanelli (BTZ) black hole \cite{Kraus:2006wn}. The torus partition function is well-known to have the Hawking--Page phase transition \cite{Hawking:1982dh} (related to the confinement--deconfinement transition in the CFT \cite{Witten:1998zw}):
\begin{equation}
    \log Z_{\text{holographic}}(\tau) = \begin{cases}
        - \frac{i \pi}{6} c \tau, \ |\tau| > 1 \ \text{(thermal AdS)}  \\
        \frac{i \pi}{6} \frac{c}{\tau}, \  |\tau|<1 \ \text{(BTZ black hole).}
    \end{cases}
\end{equation}
This implies that $Z_n$ has a phase transition at $\tau=i/n$. \figref{fig:phase} (orange) shows the corresponding critical cross ratio $x$ as a function of $n$. 

From the bulk saddle, we can focus on a Cauchy slice that includes a single copy of the subregions $A,B,C$ in order to compare with the RTN picture in \figref{fig:EWCS}. This comparison is depicted in \figref{fig:torus_slice}. The picture illustrates how the conical defect of opening angle $\frac{2\pi}{n}$ is a squeezed (expanded) entanglement wedge cross section for $n>1$ ($n<1$). We clearly see that the pictures look similar, although there is a quantitative difference between the RTN and AdS/CFT as expected from the fact that the RTN lacks backreaction. This difference is emphasized for $n<\frac{1}{2}$, where the analytic continuation for the RTN fails whereas we see that the gravitational description continues to make sense all the way to $n=0$.

\section{Conclusion}
In this paper we draw a direct link between the CCNR and $(2,n)$-R\'enyi reflected entropy. We evaluated this quantity for random tensor networks and holographic CFTs. In both cases, the CCNR has a phase transition which, in the holographic case, is essentially the Hawking--Page phase transition between thermal $AdS$ and BTZ black hole. We demonstrated explicit examples where one can vividly see the presence of a squeezed/expanded cross section. We now comment on some potential future directions.

\paragraph{Symmetry--resolved CCNR}
We can easily consider a charged BTZ black hole in the bulk. What does it correspond to at the boundary? The answer is that it corresponds to the symmetry--resolved \cite{Belin:2013uta,Goldstein:2017bua,Xavier:2018kqb,Casini:2019kex,Milekhin:2021lmq,Zhao:2020qmn} version of CCNR. Symmetry--resolved quantities arise whenever a density matrix can be decomposed into blocks (for example different representations). Then one can assign different weights to these blocks. Previously, symmetry--resolution for the reflected entropy was addressed in Ref.~\cite{Bueno:2020vnx}. The symmetry--resolution is based on the following general statement: if the state $\ket{\psi}$ is invariant under a certain global symmetry, $e^{i \mu Q} \ket{\psi}=\ket{\psi}$ and that symmetry has a split property\footnote{In simple terms it means that the charge $Q$ is the integral of a local quantity.}, then the reduced density matrix $\rho_{AB}$ on $AB$ is also invariant:
\begin{equation}
    e^{i \mu Q_{AB}} \rho_{AB} = \rho_{AB} e^{i \mu Q_{AB}}, \ Q_{AB} = Q_A + Q_B.
\end{equation}
In terms of the canonical purification, it means that the state $\ket{\sqrt{\rho_{AB}}}$ of 
$ABA^* B^*$ is invariant under the following action:
\begin{equation}
   e^{i\mu Q_{AB} - i\mu Q_{A^* B^*} }
   \ket{\sqrt{\rho_{AB}}} = 
\ket{\sqrt{\rho_{AB}}}. 
\end{equation}
Tracing out $BB^*$ implies that $\rho_{A A^*}$ and $R R^\dagger$ are invariant under $e^{i \mu Q_A - i\mu Q_{A^*}}$. Now we can introduce the symmetry--resolved R\'enyi version of CCNR (cf. \Eqref{eq:renyi}):
\begin{equation}
    Z_n(\mu) = \Tr \[ (R R^\dagger)^n e^{i \mu Q_A - i\mu Q_{A^*}} \].
\end{equation}
It it not difficult to see that once we glue all the replicas together we will have an extra Wilson loop wrapping one of the torus cycles. Holographically, it corresponds to non-zero black hole chemical potential. Again, we would expect two geometries (either a black hole or AdS) leading to a phase transition. Similarly, we can introduce a chemical potential for rotations, which is angular momentum. If the angular momentum is imaginary, other $SL(2,\mathbb{Z})$ black holes may become dominant \cite{Dijkgraaf:2000fq}. Interestingly, the above construction introduces an electric Wilson loop around only one of the torus cycles. In the path integral language it is straightforward to introduce a Wilson loop around the other cycle. However, its operational meaning is not clear, as the corresponding charge operator acts on a region outside $A A^*$. We leave this question for future work. 

\paragraph{Open Questions}
  Let us conclude with a list of other open questions:
  \begin{itemize}
      \item The CFT answer is well-defined for $n<1/2$, whereas the naive RTN computation reveals a square-root branch-cut below $n=1/2$. It would be interesting to understand the nature of this branch-cut in RTN. Moreover, it would be interesting to understand the validity of our proposed analytic continuation in other holographic settings (especially for $n<1$).
    \item It would be illustrative to see how the full CFT answer emerges in RTN once we sum over all possible fixed-area states. This may help understand better tensor network models of holography (for instance, see Refs.~\cite{Cheng:2022ori,negativity}).
    \item For $n=1/2$, $Z_n$ is equal to CCNR, whereas for $n=1$ it yields the purity of the state. It would be interesting to understand the interpretation of $Z_n$ for more general $n$.
    \item It is well known that thermal AdS and BTZ are not the only solutions of three-dimensional gravity. There is a whole family parametrized by the group $SL(2,\mathbb{Z})$ \cite{Dijkgraaf:2000fq,Kraus:2006wn}. In our case of pure imaginary $\tau$ all other saddles are subdominant. Nevertheless, it would be interesting to see if they have some relevance here, especially near the phase transition. 
  \end{itemize}

\acknowledgments
We would like to thank Eugenia~Colafranceschi, Xi~Dong, Sean~McBride and Amir~Tajdini for comments.
PR is supported in part by a grant from the Simons Foundation, and by funds from UCSB. This material is based upon work supported by the Air Force Office of Scientific Research under award number FA9550-19-1-0360.
 It was also supported in part by
funds from the University of California. 

\appendix

\addcontentsline{toc}{section}{References}
\bibliographystyle{JHEP}
\bibliography{main}

\providecommand{\href}[2]{#2}\begingroup\raggedright\begin{thebibliography}{10}

\bibitem{Yin:2022toc}
C.~Yin and Z.~Liu, \emph{{Universal entanglement and correlation measure in two-dimensional conformal field theory}},  \href{https://arxiv.org/abs/2211.11952}{{\ttfamily 2211.11952}}.

\bibitem{2002quant.ph..5017C}
K.~{Chen} and L.-A.~{Wu}, \emph{{A matrix realignment method for recognizing entanglement}}, {\emph{arXiv e-prints} (2002) quant} [\href{https://arxiv.org/abs/quant-ph/0205017}{{\ttfamily quant-ph/0205017}}].

\bibitem{2003JPhA...36.5825R}
O.~{Rudolph}, \emph{{On the cross norm criterion for separability}}, \href{https://doi.org/10.1088/0305-4470/36/21/311}{\emph{Journal of Physics A Mathematical General} {\bfseries 36} (2003) 5825} [\href{https://arxiv.org/abs/quant-ph/0202121}{{\ttfamily quant-ph/0202121}}].

\bibitem{Rudolph_2005}
O.~Rudolph, \emph{Further results on the cross norm criterion for separability}, \href{https://doi.org/10.1007/s11128-005-5664-1}{\emph{Quantum Information Processing} {\bfseries 4} (2005) 219–239}.

\bibitem{peres1996separability}
A.~Peres, \emph{Separability criterion for density matrices}, {\emph{Physical Review Letters} {\bfseries 77} (1996) 1413}.

\bibitem{horodecki2001separability}
M.~Horodecki, P.~Horodecki and R.~Horodecki, \emph{Separability of n-particle mixed states: necessary and sufficient conditions in terms of linear maps}, {\emph{Physics Letters A} {\bfseries 283} (2001) 1}.

\bibitem{Liu_2022}
Z.~Liu, Y.~Tang, H.~Dai, P.~Liu, S.~Chen and X.~Ma, \emph{Detecting entanglement in quantum many-body systems via permutation moments}, \href{https://doi.org/10.1103/physrevlett.129.260501}{\emph{Phys. Rev. Lett.} {\bfseries 129} (2022) 260501} [\href{https://arxiv.org/abs/2203.08391}{{\ttfamily 2203.08391}}].

\bibitem{Dutta:2019gen}
S.~Dutta and T.~Faulkner, \emph{{A canonical purification for the entanglement wedge cross-section}}, \href{https://doi.org/10.1007/JHEP03(2021)178}{\emph{JHEP} {\bfseries 03} (2021) 178} [\href{https://arxiv.org/abs/1905.00577}{{\ttfamily 1905.00577}}].

\bibitem{Dubail_2017}
J.~Dubail, \emph{Entanglement scaling of operators: a conformal field theory approach, with a glimpse of simulability of long-time dynamics in 1+1$d$}, \href{https://doi.org/10.1088/1751-8121/aa6f38}{\emph{J. Phys. A: Math. Theor.} {\bfseries 50} (2017) 234001} [\href{https://arxiv.org/abs/1612.08630}{{\ttfamily 1612.08630}}].

\bibitem{Wang:2019ued}
H.~Wang and T.~Zhou, \emph{{Barrier from chaos: operator entanglement dynamics of the reduced density matrix}}, \href{https://doi.org/10.1007/JHEP12(2019)020}{\emph{JHEP} {\bfseries 12} (2019) 020} [\href{https://arxiv.org/abs/1907.09581}{{\ttfamily 1907.09581}}].

\bibitem{Dong_2021}
X.~Dong, X.-L.~Qi and M.~Walter, \emph{Holographic entanglement negativity and replica symmetry breaking}, \href{https://doi.org/10.1007/jhep06(2021)024}{\emph{Journal of High Energy Physics} {\bfseries 2021} (2021) }.

\bibitem{dong2024entanglementnegativityreplicasymmetry}
X.~Dong, J.~Kudler-Flam and P.~Rath, \emph{Entanglement negativity and replica symmetry breaking in general holographic states},  \href{https://arxiv.org/abs/2409.13009}{{\ttfamily 2409.13009}}.

\bibitem{Akers:2021pvd}
C.~Akers, T.~Faulkner, S.~Lin and P.~Rath, \emph{{Reflected entropy in random tensor networks}}, \href{https://doi.org/10.1007/JHEP05(2022)162}{\emph{JHEP} {\bfseries 05} (2022) 162} [\href{https://arxiv.org/abs/2112.09122}{{\ttfamily 2112.09122}}].

\bibitem{Akers:2022max}
C.~Akers, T.~Faulkner, S.~Lin and P.~Rath, \emph{{The Page curve for reflected entropy}}, \href{https://doi.org/10.1007/JHEP06(2022)089}{\emph{JHEP} {\bfseries 06} (2022) 089} [\href{https://arxiv.org/abs/2201.11730}{{\ttfamily 2201.11730}}].

\bibitem{Akers:2022zxr}
C.~Akers, T.~Faulkner, S.~Lin and P.~Rath, \emph{{Reflected entropy in random tensor networks II: a topological index from the canonical purification}},  \href{https://arxiv.org/abs/2210.15006}{{\ttfamily 2210.15006}}.

\bibitem{Akers:2024pgq}
C.~Akers, T.~Faulkner, S.~Lin and P.~Rath, \emph{{Reflected entropy in random tensor networks III: triway cuts}},  \href{https://arxiv.org/abs/2409.17218}{{\ttfamily 2409.17218}}.

\bibitem{Lewkowycz:2013nqa}
A.~Lewkowycz and J.~Maldacena, \emph{{Generalized gravitational entropy}}, \href{https://doi.org/10.1007/JHEP08(2013)090}{\emph{JHEP} {\bfseries 08} (2013) 090} [\href{https://arxiv.org/abs/1304.4926}{{\ttfamily 1304.4926}}].

\bibitem{Bueno:2020fle}
P.~Bueno and H.~Casini, \emph{{Reflected entropy for free scalars}}, \href{https://doi.org/10.1007/JHEP11(2020)148}{\emph{JHEP} {\bfseries 11} (2020) 148} [\href{https://arxiv.org/abs/2008.11373}{{\ttfamily 2008.11373}}].

\bibitem{Dutta:2022kge}
S.~Dutta, T.~Faulkner and S.~Lin, \emph{{The Reflected Entanglement Spectrum for Free Fermions}},  \href{https://arxiv.org/abs/2211.17255}{{\ttfamily 2211.17255}}.

\bibitem{Hayden:2016cfa}
P.~Hayden, S.~Nezami, X.-L.~Qi, N.~Thomas, M.~Walter and Z.~Yang, \emph{{Holographic duality from random tensor networks}}, \href{https://doi.org/10.1007/JHEP11(2016)009}{\emph{JHEP} {\bfseries 11} (2016) 009} [\href{https://arxiv.org/abs/1601.01694}{{\ttfamily 1601.01694}}].

\bibitem{2020PhRvL.125x1602N}
S.~{Nezami} and M.~{Walter}, \emph{{Multipartite Entanglement in Stabilizer Tensor Networks}}, \href{https://doi.org/10.1103/PhysRevLett.125.241602}{\emph{\prl} {\bfseries 125} (2020) 241602} [\href{https://arxiv.org/abs/1608.02595}{{\ttfamily 1608.02595}}].

\bibitem{Cheng:2022ori}
N.~Cheng, C.~Lancien, G.~Penington, M.~Walter and F.~Witteveen, \emph{{Random tensor networks with nontrivial links}},  \href{https://arxiv.org/abs/2206.10482}{{\ttfamily 2206.10482}}.

\bibitem{dong2023rtn}
X.~Dong, S.~McBride and W.W.~Weng, \emph{Holographic tensor networks with bulk gauge symmetries},  \href{https://arxiv.org/abs/2309.06436}{{\ttfamily 2309.06436}}.

\bibitem{Nguyen_2018}
P.~Nguyen, T.~Devakul, M.G.~Halbasch, M.P.~Zaletel and B.~Swingle, \emph{Entanglement of purification: from spin chains to holography}, \href{https://doi.org/10.1007/jhep01(2018)098}{\emph{JHEP} {\bfseries 01} (2018) 98} [\href{https://arxiv.org/abs/1709.07424}{{\ttfamily 1709.07424}}].

\bibitem{Akers:2023obn}
C.~Akers, T.~Faulkner, S.~Lin and P.~Rath, \emph{{Entanglement of purification in random tensor networks}}, \href{https://doi.org/10.1103/PhysRevD.109.L101902}{\emph{Phys. Rev. D} {\bfseries 109} (2024) L101902} [\href{https://arxiv.org/abs/2306.06163}{{\ttfamily 2306.06163}}].

\bibitem{2021JHEP...06..024D}
X.~{Dong}, X.-L.~{Qi} and M.~{Walter}, \emph{{Holographic entanglement negativity and replica symmetry breaking}}, \href{https://doi.org/10.1007/JHEP06(2021)024}{\emph{Journal of High Energy Physics} {\bfseries 2021} (2021) 24} [\href{https://arxiv.org/abs/2101.11029}{{\ttfamily 2101.11029}}].

\bibitem{2022JHEP...02..076K}
J.~{Kudler-Flam}, V.~{Narovlansky} and S.~{Ryu}, \emph{{Negativity spectra in random tensor networks and holography}}, \href{https://doi.org/10.1007/JHEP02(2022)076}{\emph{JHEP} {\bfseries 02} (2022) 76} [\href{https://arxiv.org/abs/2109.02649}{{\ttfamily 2109.02649}}].

\bibitem{Dong:2021oad}
X.~Dong, S.~McBride and W.W.~Weng, \emph{{Replica wormholes and holographic entanglement negativity}}, \href{https://doi.org/10.1007/JHEP06(2022)094}{\emph{JHEP} {\bfseries 06} (2022) 094} [\href{https://arxiv.org/abs/2110.11947}{{\ttfamily 2110.11947}}].

\bibitem{Gadde:2022cqi}
A.~Gadde, V.~Krishna and T.~Sharma, \emph{{New multipartite entanglement measure and its holographic dual}}, \href{https://doi.org/10.1103/PhysRevD.106.126001}{\emph{Phys. Rev. D} {\bfseries 106} (2022) 126001} [\href{https://arxiv.org/abs/2206.09723}{{\ttfamily 2206.09723}}].

\bibitem{Penington:2022dhr}
G.~Penington, M.~Walter and F.~Witteveen, \emph{{Fun with replicas: tripartitions in tensor networks and gravity}},  \href{https://arxiv.org/abs/2211.16045}{{\ttfamily 2211.16045}}.

\bibitem{Gadde:2024taa}
A.~Gadde, J.~Harper and V.~Krishna, \emph{{Multi-invariants and Bulk Replica Symmetry}},  \href{https://arxiv.org/abs/2411.00935}{{\ttfamily 2411.00935}}.

\bibitem{Aubrun_2012}
G.~Aubrun and I.~Nechita, \emph{Realigning random states}, \href{https://doi.org/10.1063/1.4759115}{\emph{J. Math. Phys} {\bfseries 53} (2012) 102210}.

\bibitem{Headrick:2010zt}
M.~Headrick, \emph{{Entanglement Renyi entropies in holographic theories}}, \href{https://doi.org/10.1103/PhysRevD.82.126010}{\emph{Phys. Rev. D} {\bfseries 82} (2010) 126010} [\href{https://arxiv.org/abs/1006.0047}{{\ttfamily 1006.0047}}].

\bibitem{Dong:2023bfy}
X.~Dong, J.~Kudler-Flam and P.~Rath, \emph{{A modified cosmic brane proposal for holographic Renyi entropy}}, \href{https://doi.org/10.1007/JHEP06(2024)120}{\emph{JHEP} {\bfseries 06} (2024) 120} [\href{https://arxiv.org/abs/2312.04625}{{\ttfamily 2312.04625}}].

\bibitem{2019JHEP...10..240D}
X.~{Dong}, D.~{Harlow} and D.~{Marolf}, \emph{{Flat entanglement spectra in fixed-area states of quantum gravity}}, \href{https://doi.org/10.1007/JHEP10(2019)240}{\emph{Journal of High Energy Physics} {\bfseries 2019} (2019) 240} [\href{https://arxiv.org/abs/1811.05382}{{\ttfamily 1811.05382}}].

\bibitem{Akers:2018fow}
C.~Akers and P.~Rath, \emph{{Holographic Renyi Entropy from Quantum Error Correction}}, \href{https://doi.org/10.1007/JHEP05(2019)052}{\emph{JHEP} {\bfseries 05} (2019) 052} [\href{https://arxiv.org/abs/1811.05171}{{\ttfamily 1811.05171}}].

\bibitem{Lunin:2002fch}
O.~Lunin and S.D.~Mathur, \emph{{Correlation functions for orbifolds of the type M$^{N}$/S$^{N}$}}, {\emph{AMS/IP Stud. Adv. Math.} {\bfseries 33} (2002) 311}.

\bibitem{Banados:1992wn}
M.~Ba\~nados, C.~Teitelboim and J.~Zanelli, \emph{{Black hole in three-dimensional spacetime}}, \href{https://doi.org/10.1103/PhysRevLett.69.1849}{\emph{Phys. Rev. Lett.} {\bfseries 69} (1992) 1849} [\href{https://arxiv.org/abs/hep-th/9204099}{{\ttfamily hep-th/9204099}}].

\bibitem{Maldacena:1998bw}
J.M.~Maldacena and A.~Strominger, \emph{{AdS$_3$ black holes and a stringy exclusion principle}}, \href{https://doi.org/10.1088/1126-6708/1998/12/005}{\emph{JHEP} {\bfseries 12} (1998) 005} [\href{https://arxiv.org/abs/hep-th/9804085}{{\ttfamily hep-th/9804085}}].

\bibitem{Dijkgraaf:2000fq}
R.~Dijkgraaf, J.M.~Maldacena, G.W.~Moore and E.P.~Verlinde, \emph{{A black hole {F}arey tail}},  \href{https://arxiv.org/abs/hep-th/0005003}{{\ttfamily hep-th/0005003}}.

\bibitem{Boer_2006}
J.d.~Boer, M.C.~Cheng, R.~Dijkgraaf, J.~Manschot and E.~Verlinde, \emph{A {F}arey tail for attractor black holes}, \href{https://doi.org/10.1088/1126-6708/2006/11/024}{\emph{JHEP} {\bfseries 11} (2006) 024} [\href{https://arxiv.org/abs/hep-th/0608059}{{\ttfamily hep-th/0608059}}].

\bibitem{Kraus:2006wn}
P.~Kraus, \emph{{Lectures on black holes and the AdS(3) / CFT(2) correspondence}}, {\emph{Lect. Notes Phys.} {\bfseries 755} (2008) 193} [\href{https://arxiv.org/abs/hep-th/0609074}{{\ttfamily hep-th/0609074}}].

\bibitem{Hawking:1982dh}
S.W.~Hawking and D.N.~Page, \emph{{Thermodynamics of Black Holes in anti-De Sitter Space}}, \href{https://doi.org/10.1007/BF01208266}{\emph{Commun. Math. Phys.} {\bfseries 87} (1983) 577}.

\bibitem{Witten:1998zw}
E.~Witten, \emph{{Anti-de Sitter space, thermal phase transition, and confinement in gauge theories}}, \href{https://doi.org/10.4310/ATMP.1998.v2.n3.a3}{\emph{Adv. Theor. Math. Phys.} {\bfseries 2} (1998) 505} [\href{https://arxiv.org/abs/hep-th/9803131}{{\ttfamily hep-th/9803131}}].

\bibitem{Belin:2013uta}
A.~Belin, L.-Y.~Hung, A.~Maloney, S.~Matsuura, R.C.~Myers and T.~Sierens, \emph{{Holographic Charged Renyi Entropies}}, \href{https://doi.org/10.1007/JHEP12(2013)059}{\emph{JHEP} {\bfseries 12} (2013) 059} [\href{https://arxiv.org/abs/1310.4180}{{\ttfamily 1310.4180}}].

\bibitem{Goldstein:2017bua}
M.~Goldstein and E.~Sela, \emph{{Symmetry-resolved entanglement in many-body systems}}, \href{https://doi.org/10.1103/PhysRevLett.120.200602}{\emph{Phys. Rev. Lett.} {\bfseries 120} (2018) 200602} [\href{https://arxiv.org/abs/1711.09418}{{\ttfamily 1711.09418}}].

\bibitem{Xavier:2018kqb}
J.C.~Xavier, F.C.~Alcaraz and G.~Sierra, \emph{{Equipartition of the entanglement entropy}}, \href{https://doi.org/10.1103/PhysRevB.98.041106}{\emph{Phys. Rev. B} {\bfseries 98} (2018) 041106} [\href{https://arxiv.org/abs/1804.06357}{{\ttfamily 1804.06357}}].

\bibitem{Casini:2019kex}
H.~Casini, M.~Huerta, J.M.~Mag\'an and D.~Pontello, \emph{{Entanglement entropy and superselection sectors. Part I. Global symmetries}}, \href{https://doi.org/10.1007/JHEP02(2020)014}{\emph{JHEP} {\bfseries 02} (2020) 014} [\href{https://arxiv.org/abs/1905.10487}{{\ttfamily 1905.10487}}].

\bibitem{Milekhin:2021lmq}
A.~Milekhin and A.~Tajdini, \emph{{Charge fluctuation entropy of Hawking radiation: a replica-free way to find large entropy}},  \href{https://arxiv.org/abs/2109.03841}{{\ttfamily 2109.03841}}.

\bibitem{Zhao:2020qmn}
S.~Zhao, C.~Northe and R.~Meyer, \emph{{Symmetry-resolved entanglement in AdS$_{3}$/CFT$_{2}$ coupled to U(1) Chern-Simons theory}}, \href{https://doi.org/10.1007/JHEP07(2021)030}{\emph{JHEP} {\bfseries 07} (2021) 030} [\href{https://arxiv.org/abs/2012.11274}{{\ttfamily 2012.11274}}].

\bibitem{Bueno:2020vnx}
P.~Bueno and H.~Casini, \emph{{Reflected entropy, symmetries and free fermions}}, \href{https://doi.org/10.1007/JHEP05(2020)103}{\emph{JHEP} {\bfseries 05} (2020) 103} [\href{https://arxiv.org/abs/2003.09546}{{\ttfamily 2003.09546}}].

\end{thebibliography}\endgroup


\end{document}